\documentclass{article}

\usepackage{color}\usepackage{epsfig}
\usepackage{amssymb}\usepackage{bm}\usepackage{graphicx}

\begin{document}
\title{Sandpiles and superconductors:
dual variational formulations for critical-state problems}
%
\author{John W. Barrett$^1$ and \underline{Leonid Prigozhin}$^2$\\
\footnotesize $^1$Dept.\ of
Mathematics, Imperial College, London SW7 2AZ, UK.\\
\footnotesize $^2$Blaustein Inst.\ for Desert Res., Ben-Gurion
Univ., Sede Boqer Campus 84990, Israel.
}
\date{}

\maketitle

\smallskip

\begin{abstract}Similar evolutionary variational inequalities appear
as convenient formulations for continuous  models for sandpile
growth, magnetization of type-II superconductors, and evolution of
some other dissipative systems characterized by the multiplicity
of metastable states, long-range interactions, avalanches, and
hysteresis. The origin of this similarity is that these are
quasistationary models of equilibrium in which the  multiplicity
of metastable states is a consequence of a unilateral condition of
equilibrium (critical-state constraint). Existing variational
formulations for critical-state models of sandpiles and
superconductors are convenient for modelling only the ``primary"
variables  (evolving pile shape and magnetic field,
respectively). The conjugate variables (the surface sand flux
and the electric field) are also of interest in various
applications. Here we derive dual variational formulations,
similar to mixed variational inequalities in plasticity, for
the sandpile and superconductor models. These formulations are used in
numerical simulations and allow us to approximate simultaneously both the
primary and dual variables.
\end{abstract}

\newpage
Sandpiles and type-II superconductors are examples of spatially
extended open dissipative systems which have infinitely many
metastable states but, driven by the external forces, tend to
organize themselves into a marginally stable ``critical state" and
are then able to demonstrate almost instantaneous long-range
interactions. The evolution of such systems is often accompanied
by sudden collapses, like sandpile avalanches, and hysteresis.
Although these are dissipative systems of a different nature,
their continuous models are equivalent to similar variational (or
quasivariational) inequalities (see \cite{PhysD} and the
references therein). The origin of this similarity is that these
models are quasistationary models of equilibrium and the
multiplicity of metastable states is a consequence of a unilateral
equilibrium condition. The rate with which such a system adjusts
itself to the changing external conditions is determined
implicitly and appears in the model as a Lagrange multiplier.
Typically, the multiplier is eliminated in transition to a
variational formulation written in terms of a ``primary" variable
(surface of a sandpile, magnetic field in a superconductor, stress
tensor in elastoplasticity, etc.) In many situations, however, the
Lagrange multiplier or, equivalently, a ``dual" variable (sand
flux upon the pile surface, electric field, and strain tensor,
respectively) also has to be found. We present variational
formulations, for both the sandpile and the superconductivity
problem, written for the dual variables. Only the simplest version
of each problem is considered.

\underline{\it Sandpiles} -- Let sand be poured out onto a rigid
support surface, $y=h_0(x)$, given in a domain $\Omega\subset
\mathbb{R}^2$ with boundary $\partial \Omega$. If the support boundary is open, a model for pile
surface evolution can be written as

\centerline{$
\partial_t h+\nabla\cdot \bm{q}=w,\ \ \ h|_{t=0}=h_0,\ \ \
\left. h\right|_{\partial\Omega}=\left.
h_0\right|_{\partial\Omega},$}

\noindent where $h$ is the pile surface, $w\ge 0$ is the given
source density,  $\bm{q}$ is the horizontal projection of the flux
of sand pouring down the pile surface. If the support has no
slopes steeper than the sand angle of repose, $|\nabla h_0|\le
\gamma=\tan\alpha$, the simplest {constitutive relations} for this
model  read: (i) the flux $\bm{q}$ is directed towards the
steepest decent of the surface, (ii) the surface slope cannot
exceed the critical angle $\alpha$, and (iii) the flux is zero
upon subcritical slopes. Equivalently, one can write
$\bm{q}=-m\nabla h$ and show that $m(x,t)\ge 0$ is the Lagrange
multiplier related to the constraint $|\nabla h|\le \gamma$. The
model can be rewritten as a variational inequality for $h$,

\vspace{-.2cm}

\begin{equation}h(.,t)\in K:\ (\partial_t h-w,\varphi-h)\ge 0\
\ \forall \varphi\in K,\ \
h|_{t=0}=h_0,\label{vi_sand}\end{equation}

\vspace{-.2cm}

\noindent where $K=\{\varphi\in H^1(\Omega)\ :\ |\nabla
\varphi|\le \gamma\ a.e.,\ \left.
\varphi\right|_{\partial\Omega}=\left.
h_0\right|_{\partial\Omega}\}$ and $(\cdot,\cdot)$ is the standard $L^2(\Omega)$
inner product.
 Simple analytical solutions of this inequality describe
piles generated on the support $h_0\equiv 0$. For the point source
$w=\delta(x-x_0)$, a conical pile with critical slopes grows until
its base touches the domain boundary. Then there appears a runway
connecting the cone apex with the boundary and the pile growth
stops: all additional sand just follows the runway and leaves the
system. On the other hand, if $w>0$ everywhere in $\Omega$, the
final stationary shape of the pile is different: $h(x)=\gamma
\mbox{dist}(x,\partial\Omega)$. For $w\ge 0$, the general {
stationary} solution and an integral representation formula for
the corresponding Lagrange multiplier $m$, determining the surface
sand flux $\bm{q}$, have also been obtained recently (see \cite{CC}
and the references therein). Note, however, that it is not easy to
compute the Lagrange multiplier using this formula. In the
non-stationary case, determining the surface flux $\bm{q}$ remains
difficult even if  the unique solution $h$ to (\ref{vi_sand}) is
found. To compute both these variables simultaneously, we derive a
dual variational formulation of the { evolutionary} problem.

Let 
$\{h,\bm{q}\}$ satisfy the model relations (i)-(iii).
Then, for any test field $\bm{\psi}$, 
{ $\nabla h\cdot(\bm{\psi}-\bm{q})\ge -|\nabla
h||\bm{\psi}|-\nabla h\cdot\bm{q}=-|\nabla
h||\bm{\psi}|+\gamma|{\bm{q}}|\ge
-\gamma|\bm{\psi}|+\gamma|{\bm{q}}|$.} \noindent Hence, $(\nabla
h,\bm{\psi}-\bm{q})\ge\phi(\bm{q})-\phi(\bm{\psi}),$ where
$\phi(\bm{q})=\gamma\int_{\Omega}|{\bm{q}}|$. Since $(\nabla
h,\bm{\psi}-\bm{q})=\oint_{\partial\Omega}h_0\{\bm{\psi}_n-\bm{q}_n\}
-(h,\nabla\cdot\{\bm{\psi}-\bm{q}\})$, where $\bm{\psi}_n$ is the
normal component of $\bm{\psi}$ on $\partial \Omega$, we have
$\phi(\bm{\psi})-\phi(\bm{q})-(h,\nabla\cdot\{\bm{\psi}-\bm{q}\})
+\oint_{\partial\Omega}h_0\{\bm{\psi}_n-\bm{q}_n\}\ge
0.$
Noting that
$h=h_0+\int_0^tw\,dt-\nabla\cdot\{\int_0^t\bm{q}\,dt\}$ we finally
obtain
\begin{equation}\bm{q}(.,t)\in V:\ \left(\nabla\cdot\left\{\int_0^t\bm{q}\,dt\right\},
\nabla\cdot\{\bm{\psi}-\bm{q}\}\right)
+\mathcal{F}(\bm{\psi}-\bm{q})+\phi(\bm{\psi})-\phi(\bm{q})\ge 0
\label{vi2_sand}\end{equation} for any $\bm{\psi} \in V$. Here
$\mathcal{F}(\bm{u})=\oint_{\partial\Omega}h_0\bm{u}_n-(h_0+\int_0^tw\,
dt,\nabla\cdot\bm{u})$
and we define $V=\{\bm{\psi} \in [{\mathcal M}(\Omega)]^2\ :\
\nabla\cdot\bm{\psi}\in L^2(\Omega)\}$, where ${\mathcal
M}(\Omega)$ is the Banach space of bounded Radon measures.

\begin{figure}[h!]
\centering 
\includegraphics[width=3.9cm,height=3.9cm]{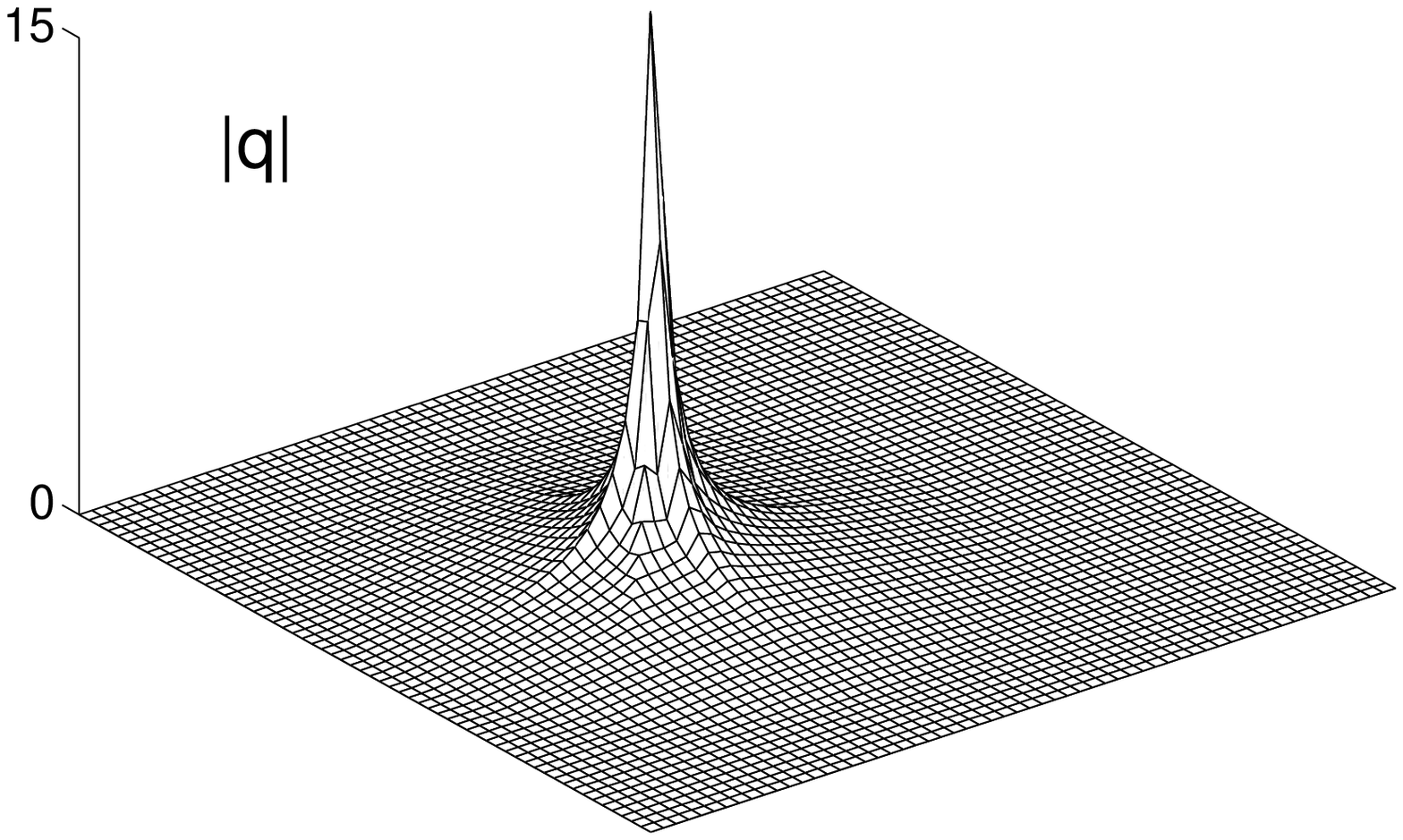}\hspace{.001cm}
\includegraphics[width=3.9cm,height=3.9cm]{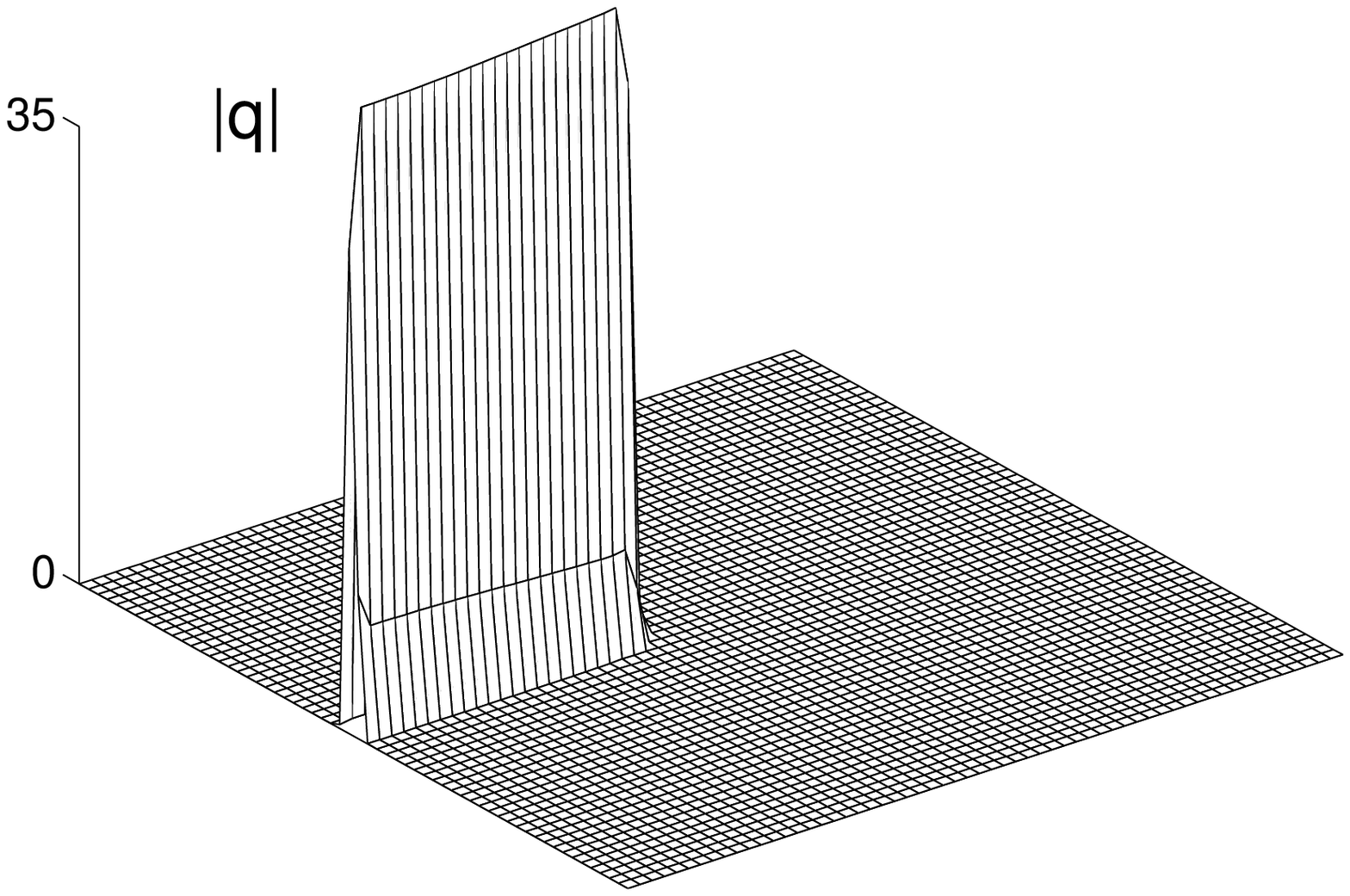}\hspace{.001cm}
\includegraphics[width=3.9cm,height=3.9cm]{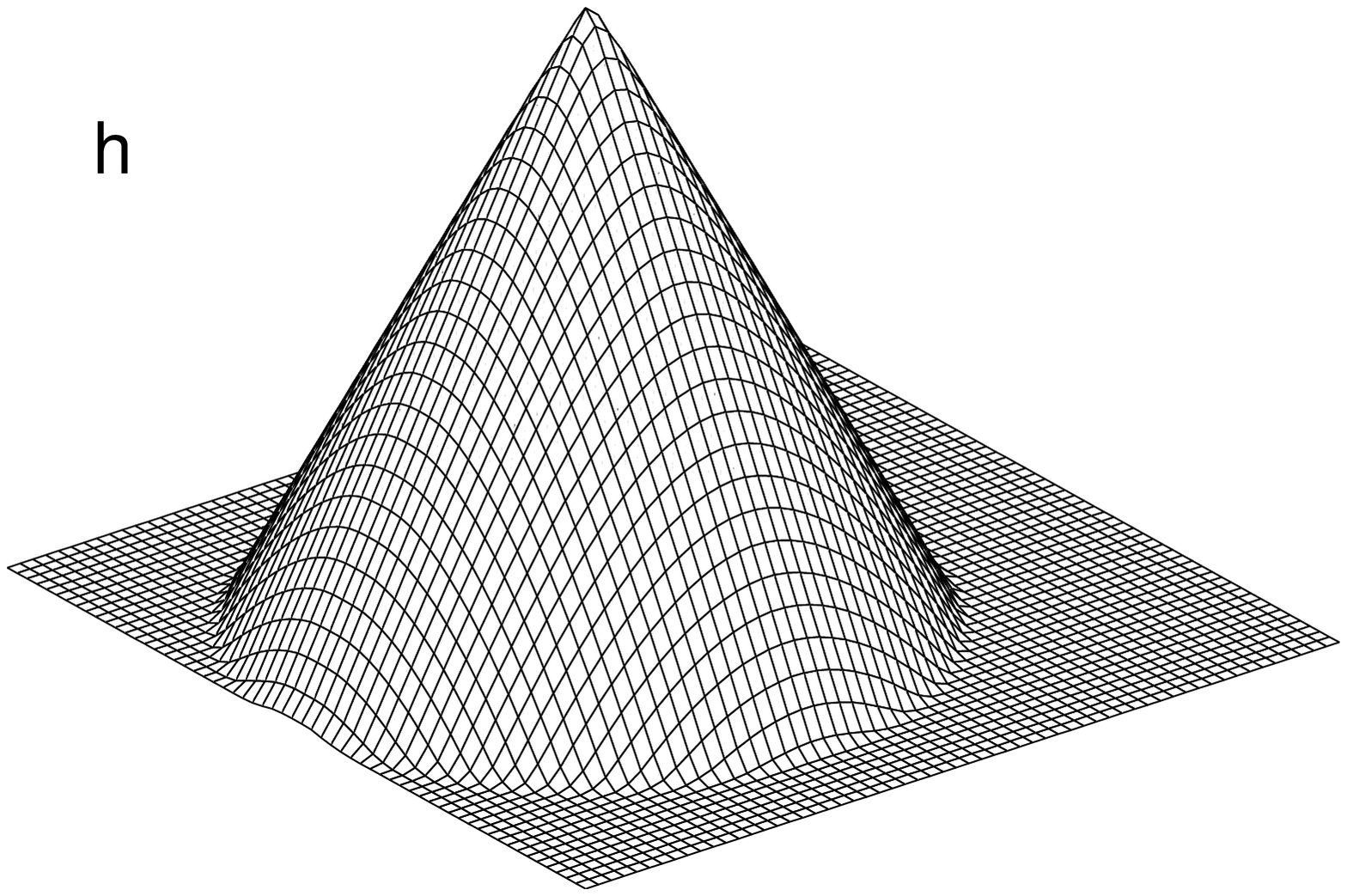}
\caption{\footnotesize Point source above square support with
$h_0= 0$. The cone grows until its base touches the support
boundary and the runway appears. Shown: 1) sand flux $|\bm{q}|$
before and after this moment; 2) the final pile shape computed as
$h=h_0+\int_0^tw\,dt-\nabla\cdot \{\int_0^t\bm{q}\,dt\}$.
\label{Fig1}}
\end{figure}

To approximate (2) 
numerically, we smoothed the
non-differentiable functional $\phi$ by introducing
$|\bm{q}|_{\varepsilon}=(|\bm{q}|^2+\varepsilon^2)^{1/2}$,
discretized the regularized equality problem in time, employed
Raviart-Thomas finite elements of lowest
order with vertex sampling  on the nonlinear term,
and solved the resulting nonlinear algebraic system
at each time level iteratively using a form of successive
over-relaxation (see Fig. \ref{Fig1} for an example of a numerical
simulation).
\vspace{-.25cm}

\underline{\it Superconductors} -- Phenomenologically, the
magnetic field penetration into type-II superconductors can be
understood as a nonlinear eddy current problem. Let a long
cylindrical superconductor with a simply connected cross-section
$\Omega$ be placed into a non-stationary uniform external magnetic
field $\bm{h}_e(t)$ parallel to the cylindrical generators.
According to Faraday's law, time variations of this field induce in
a conductor an electric field $\bm{e}$ leading to a current
$\bm{j}$ parallel to the cross-section plane; this current induces
a magnetic field $\bm{h}(x,t)$ parallel to $\bm{h}_e$. Omitting the
displacement current in Maxwell's equations and scaling the magnetic
permeability to be unity, we obtain the following model,

\centerline{$\partial_t (h+h_e)+\mbox{curl}\,\bm{e}=0,\ \
\mbox{\bf curl}\, h=\bm{j},\ \ h|_{t=0}=h_0(x),\ \ h|_{\partial
\Omega}=0,\label{sc}$}

 \noindent where
$\mbox{curl}\,\bm{u}=\partial_{x_1}u_2-\partial_{x_2}u_1$ and
$\mbox{\bf curl}\,u=(\partial_{x_2}u,-\partial_{x_1}u)$. Instead
of the usual Ohm law, a multivalued   current-voltage relation
(the Bean model) is often employed for type-II superconductors. It
is postulated that ({i}) the electric field $\bm{e}$
and the current density $\bm{j}$ have the same direction, ({ii}) the
current density cannot exceed some critical value, $j_c$, and
({iii}) if the current is subcritical, the electric field is zero.
One can write $\bm{e}=\rho\bm{j}$ and show that the effective
resistivity, $\rho(x,t)\ge 0$, is the Lagrange multiplier related to
current density constraint $|\bm{j}|\le j_c$. Using conditions
({i})-({iii}) we can eliminate the electric field from the model.
This yields the variational inequality,

\centerline{$h (.,t)\in K:\ (\partial_t\{h+h_e\},\varphi-h)\ge 0\
\ \forall \varphi\in K,\ \ h|_{t=0}=h_0,$}

 \noindent  where $K=\{\varphi\in H_0^1(\Omega)\ :\
|\nabla \varphi|\le \gamma\ a.e.\}$. This inequality for $h$ can
be approximated numerically and often even solved analytically. However,
computing the electric field $\bm{e}$ may be difficult
 \cite{BL}.
As for the sandpile model, a dual variational formulation  can be
derived to find both variables simultaneously:
\begin{equation}
\bm{e}(.,t)\in W:\
\left(\mbox{curl}\,\left\{\int_0^t\bm{e}\,dt\right\},\mbox{curl}\,
\{\bm{\psi}-\bm{e}\}\right)+{\mathcal
F}(\bm{\psi}-\bm{e})+\phi(\bm{\psi})-\phi(\bm{e})\ge
0\label{vi_sc}\end{equation} for any $\bm{\psi}\in W$, where $W=\{
\bm{\psi}\in [{\mathcal M}(\Omega)]^2\ :\ \mbox{curl}
\,\bm{\psi}\in L^2(\Omega)\}$,
$\phi(\bm{u})=\int_{\Omega}j_c|\bm{u}|$, and ${\mathcal
F}(\bm{u})=(h_e(t)-h_e(0)-h_0,\mbox{curl} \,\bm{u})$. The primary
variable, $h$, is found as

\centerline{$h=h_0+h_e(0)-h_e(t)-\mbox{curl}
\,\{\int_0^t\bm{e}\,dt\}$.}

\begin{figure}[h!]
\centering
\includegraphics[width=3.8cm,height=3.8cm]{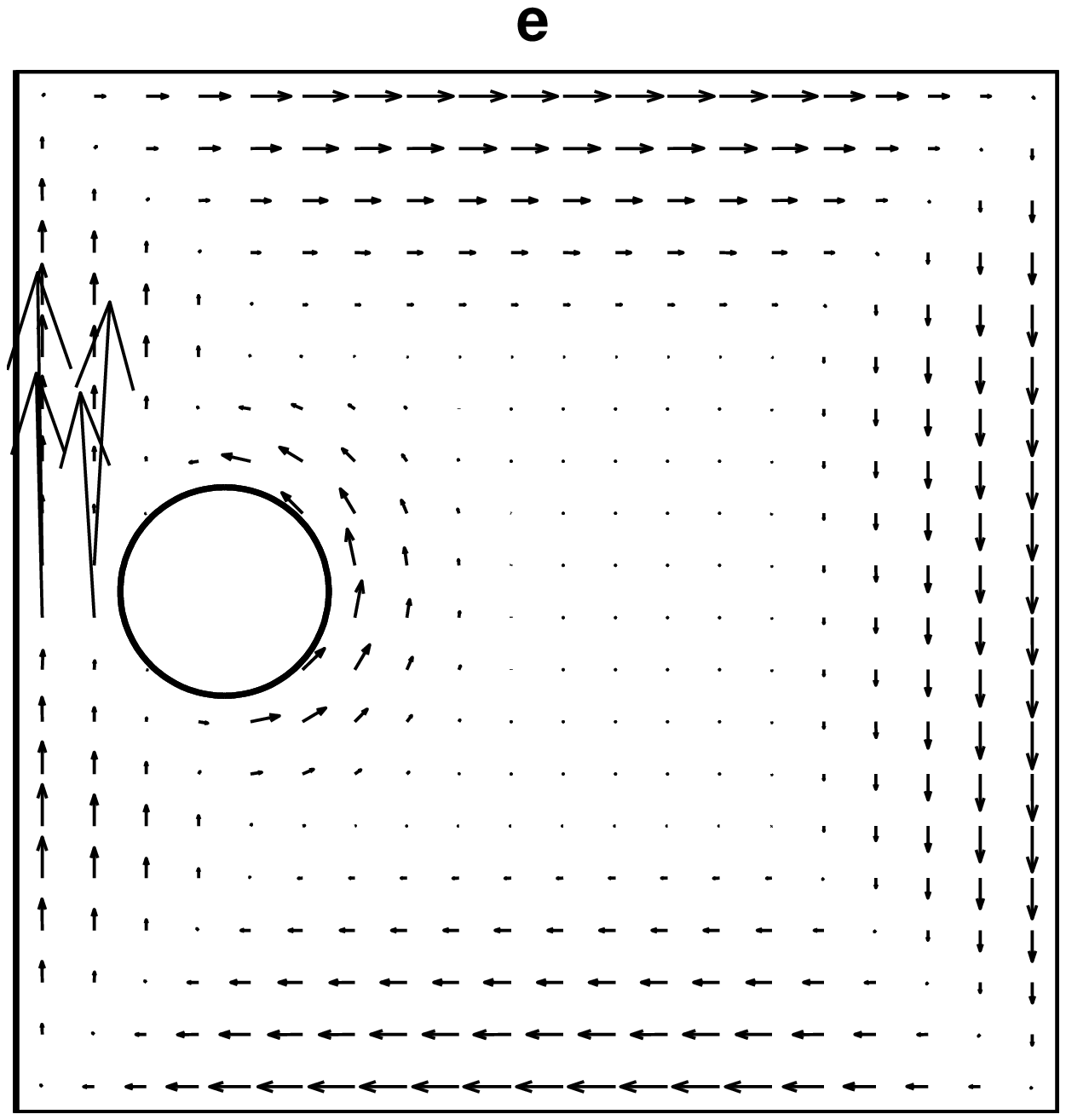}\hspace{.05cm}
\includegraphics[width=3.8cm,height=3.8cm]{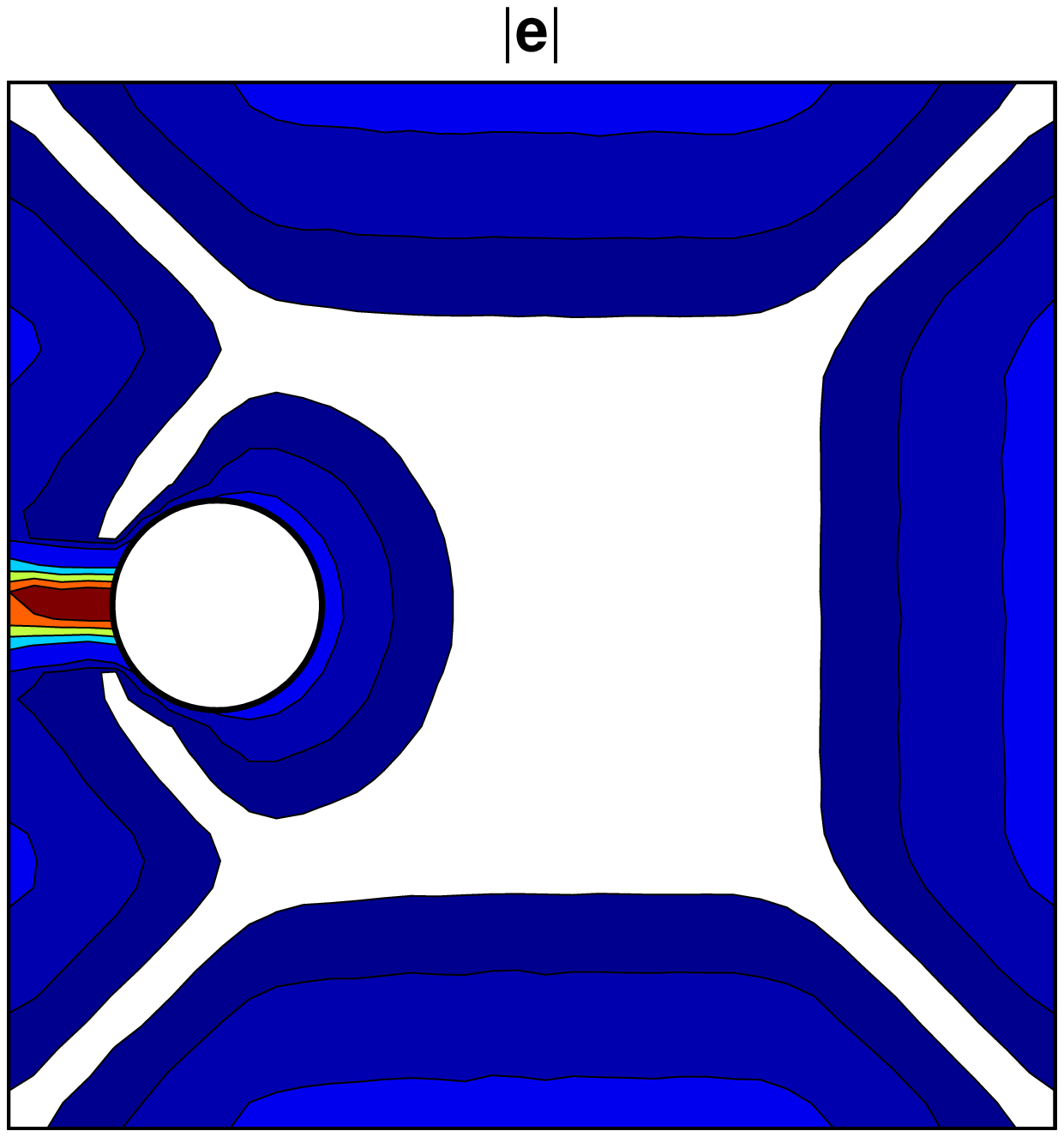}\hspace{.05cm}
\includegraphics[width=3.8cm,height=3.8cm]{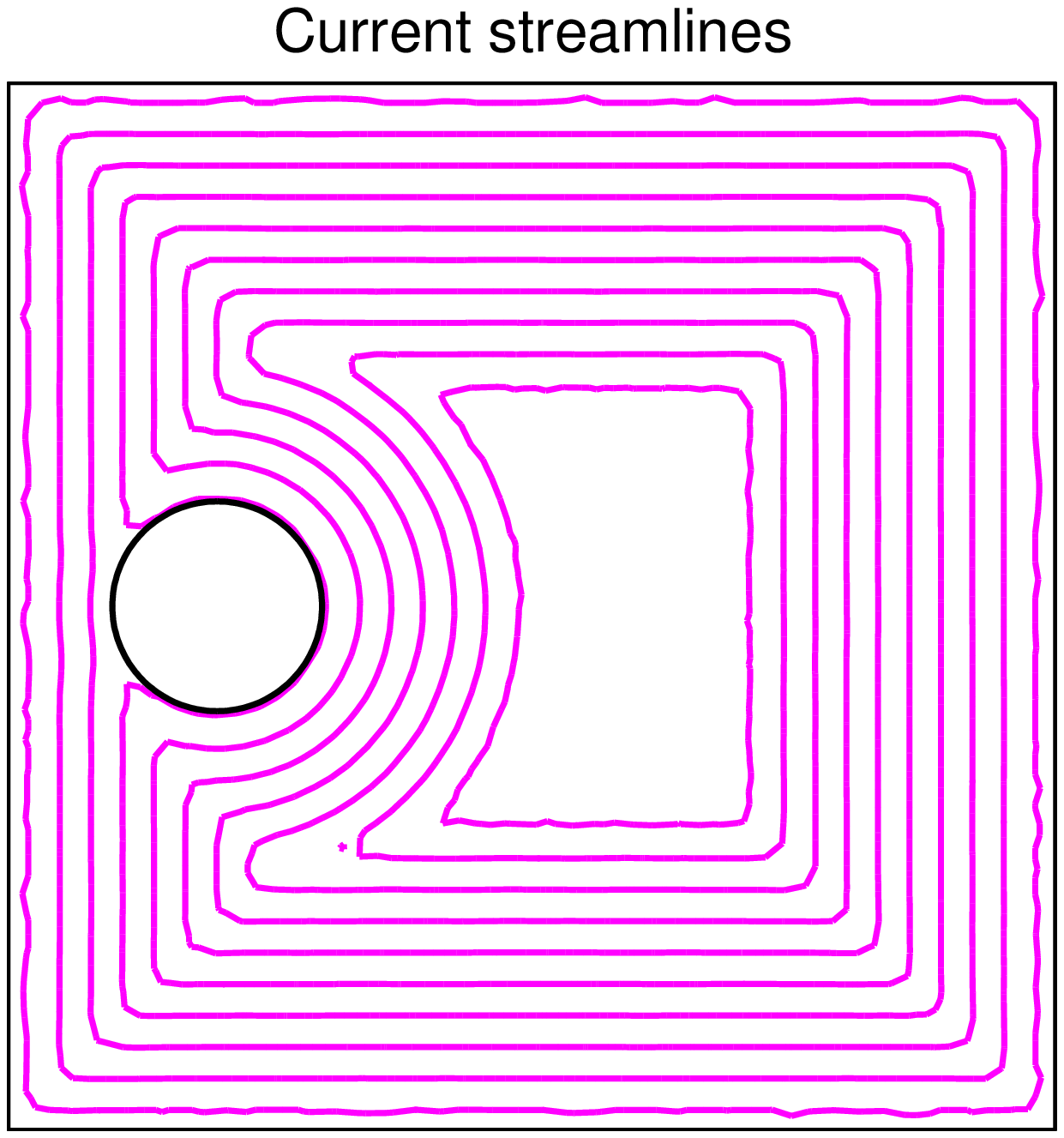}
\caption{\footnotesize Cylindrical superconductor in a growing
field (square cross-section with a hole, zero initial state).
Shown at the same moment in  time: the electric field $\bm{e}$, level
contours of $|\bm{e}|$, and the current streamlines plotted as
levels of $h$. Note the ``runway" (red region in the $|\bm{e}|$
contour plot) through which the magnetic field penetrates the hole
and where the electric field is the strongest. \label{Fig2}}
\end{figure}

\noindent  The simple transformation
$R:\bm{e}=(e_1,e_2)\rightarrow(e_2,-e_1)$ maps $W$ to $V$ and
enables us to use the same 
Raviart-Thomas finite element as in
the previous case. To model the magnetization of a superconductor with a
multiply connected cross-section (Fig. \ref{Fig2}) we ``filled"
the hole and set $j_c=0$ there.

\end{document}